\documentclass[12pt,preprint]{aastex}
\usepackage{emulateapj5,apjfonts,psfig}

\newcommand{\Msun}{M_\odot}
\newcommand{\mbh}{M_\bullet}

\newcommand{\vdm}{van~der~Marel}

\newcommand {\et}{{\it et~al.}\ }

\begin{document}

\shorttitle{Galaxy Models}

\submitted{ApJL}

\title{Black Hole Mass Determinations From Orbit Superposition Models
are Reliable}

\author{
Douglas Richstone\altaffilmark{1},
Karl Gebhardt\altaffilmark{2},
Monique Aller \altaffilmark{1},
Ralf Bender\altaffilmark{3},
Gary Bower\altaffilmark{4},
Alan Dressler\altaffilmark{5},
S.M.~Faber\altaffilmark{6},
Alexei V. Filippenko\altaffilmark{7},
Richard Green\altaffilmark{8},
Luis C.  Ho\altaffilmark{5},
John Kormendy\altaffilmark{2},
Tod R. Lauer\altaffilmark{8},
John Magorrian\altaffilmark{9},
Jason Pinkney\altaffilmark{10},
Christos Siopis\altaffilmark{1}, and Scott~Tremaine\altaffilmark{11}}

\altaffiltext{1}{Dept. of Astronomy, Dennison Bldg., Univ. of
Michigan, Ann Arbor 48109; dor@umich.edu, maller@umich.edu,
siopis@umich.edu}

\altaffiltext{2}{Department of Astronomy, University of Texas, Austin,
Texas 78712; gebhardt@astro.as.utexas.edu, kormendy@astro.as.utexas.edu}

\altaffiltext{3}{Universit\"ats-Sternwarte, Scheinerstrasse 1,
M\"unchen 81679, Germany; bender@usm.uni-muenchen.de}

\altaffiltext{4}{Computer Sciences Corporation, Space Telescope
Science Institute, 3700 San Martin Drive, Baltimore, MD 21218;
bower@stsci.edu}

\altaffiltext{5}{The Observatories of the Carnegie Institution of
Washington, 813 Santa Barbara St., Pasadena, CA 91101;
dressler@ociw.edu, lho@ociw.edu}

\altaffiltext{6}{UCO/Lick Observatories, University of California,
Santa Cruz, CA 95064; faber@ucolick.org}

\altaffiltext{7}{Department of Astronomy, University of California,
Berkeley, CA 94720-3411; alex@astro.berkeley.edu}

\altaffiltext{8}{National Optical Astronomy Observatories, P. O. Box
26732, Tucson, AZ 85726; green@noao.edu, lauer@noao.edu}

\altaffiltext{9}{Theoretical Physics, University of Oxford,
1 Keble Road, Oxford OX1 3NP,UK; j.magorrian1@physics.ox.ac.uk}

\altaffiltext{10}{Department of Physics \& Astronomy, University of
Northern Ohio, Ada, OH 45810; j-pinkney@onu.edu}

\altaffiltext{11}{Princeton University Observatory, Peyton Hall,
Princeton, NJ 08544; tremaine@astro.princeton.edu}

\begin{abstract}

We show that orbit-superposition dynamical models (Schwarzschild's
method) provide reliable estimates of nuclear black hole masses and
errors when constructed from adequate orbit libraries and kinematic
data.  We thus rebut two recent papers that argue that BH masses
obtained from this method are unreliable.  These papers claim to
demonstrate that the range of allowable BH masses derived from a given
dataset is artificially too narrow as a result of an inadequate number
of orbits in the library used to construct dynamical models.  This is
an elementary error that is easily avoided.  We describe a method to
estimate the number and nature of orbits needed for the library.  We
provide an example that shows that this prescription is adequate, in
the sense that the range of allowable BH masses is not artificially
narrowed by use of too few orbits.  This is illustrated by showing
that the $\chi^2$ versus BH-mass curve does not change beyond a
certain point as more orbits are added to the library.  At that point,
the phase-space coverage of the orbit library is good enough to
estimate the BH mass, and the $\chi^2$ profile provides a reliable
estimate of its errors.

A second point raised by critics is that kinematic data are generally
obtained with insufficient spatial resolution (compared to the BH
radius of influence) to obtain a reliable mass.  We make the
distinction between {\em unreliable} determinations and {\em
imprecise} ones.  We show that there are several different properties
of a kinematic dataset that can lead to {\em imprecise} BH
determinations (insufficient resolution among them), but none of the
attributes we have investigated leads to an unreliable determination.
In short, the degree to which the BH radius of influence is resolved
by spectroscopic observations is already reflected in the BH-mass
error envelope, and is not a hidden source of error.  The BH masses
published by our group and the Leiden group are reliable.

\end{abstract}


\section{Introduction}

There are almost twenty detections of massive black holes
(BHs) in galaxy centers that employ the technique of orbit superposition
modeling \citep{vdm98,crvdb99, cdzdm99, geb00a, geb03, capp1459,
  verolme02b}.  The orbit superposition technique is based on a method
originally invented by \citet{sch79}, who noted that the
time-averaged orbits in a stationary potential can be summed (at
finite spatial resolution) to match the mass distribution that gives
rise to the potential, thereby producing an equilibrium dynamical model.
The Schwarzschild method was first used to
analyze kinematic data for evidence of BHs in galaxy centers by
Richstone \& Tremaine (1985) for M87 and Dressler \& Richstone (1988)
for M31 and M32.  Those early models were spherically symmetric.
In modern implementations of this method, large sets of orbits are run in
a specified axisymmetric potential (based on the starlight
distribution and a central point mass) and then a non-negative linear
superposition of orbits is found that best matches the kinematic and
photometric observations.  The usual goal of this process is the
determination of only two quantities, the stellar mass-to-light
ratio $\Upsilon$ (assumed to be independent of position) and the BH
mass $\mbh$.

\citet{valluri04} (hereafter VME) and \cite{cretton04} have challenged
the reliability of BH mass determinations obtained via orbit
superposition modeling.  Cretton \& Emsellem also argue that
the uncertainties noted by VME can be dealt with via
regularization (smoothing the distribution of orbit weights).

VME raise a long list of problems and features of the
orbit-superposition analysis technique.  They focus on two main
issues: orbit insufficiency and data insufficiency.  First, they
contend that the BH mass determination is artificially narrowed
through the use of too small an orbit library and imply, but do not
show, that we have made this mistake. In addition they suggest that
the absence of a flat-bottomed $\chi^2$ profile as a function of BH
mass indicates that this problem has occurred.  Second, they suggest
that the BH mass $\mbh$ can only be determined if the radius of
influence $r_i = G\mbh/\sigma^2$ where the BH dominates the stellar
dynamics is well-resolved spectroscopically (here $\sigma$ is the
line-of-sight velocity dispersion).  In each of their main points they
have fastened on one facet of rather complex problems.

We show in \S 2 that the use of an orbit library which is too small is
an easily avoided elementary trap.  In \S 3 we refer the reader to
other work in which we show that we have adequate resolution to
determine BH masses reliably, and argue that the presence of extensive
radial kinematic coverage mitigates the uncertainties associated with
a ``not well-resolved'' radius of influence.  We also show that the
``flat-bottomed'' behavior of $\chi^2$ is a feature whose presence
depends on several features of the quality and extent of the kinematic
data, not just the size of the orbit library.

The principal result is that a $\chi^2$ profile constructed using our
method does indicate the range of acceptable BH masses in a manner
easily judged by the readers of our papers.

\section{Enough is Enough: Orbit Sufficiency}

There appear to be at least three independent axisymmetric
orbit-superposition codes, one developed by various Leiden students
starting with \vdm\ and continuing through Verolme and Cappellari,
another developed by Gebhardt and Richstone and used by our
collaboration (the ``nukers'') since 1996, and a third described by
VME.  All practitioners agree that the orbits can be sampled by
examining launch points in the three-dimensional space spanned by the
isolating integrals: the energy per unit mass $E$, the $z$-component
of the angular momentum $l_z$, and the third integral $I_3$. We
further agree that this space is finite---only bound orbits matter
($E<0$), the range of $ l_z$ at fixed energy lies between zero and the
angular momentum of a circular orbit of that energy, and the
distribution of the third integral can be sampled by launching the
orbits with zero velocity from positions spread across the zero
velocity surface at fixed $E$ and $l_z$.  The remaining question is
how densely or sparsely spaced these orbital parameters need to be
(and therefore how many orbits are required).  The answer depends in
part on the spatial resolution sought in the model, and in part on the
complexity of the orbit structure in phase space, and therefore on the
mass distribution.

Nonetheless, it is possible to estimate the number of orbits required to
represent all possibilities.  The eventual construction of the model
requires matching the (stellar) mass density in a set of bins spanning
the physical space of the (axisymmetric) model This bin set is
composed of one radial dimension (of $n_r$ bins, which are
logarithmically spaced except very near the center) and one angular
dimension (of $n_\beta$ bins, which are equally spaced in $\sin
(\beta)$ where $\beta$ is the latitude).  To fully sample the
equatorial orbits, we follow orbits with apocenter and pericenter in
all possible pairs of radial bins; this requires roughly $n_r^2/2$
orbits.  We then assign $E$ and $l_z$ from the apo-peri sampling of
the equatorial orbits and then use the same $E$ and $l_z$ for
non-equatorial orbits.  For each $E$ and $l_z$ pair, we drop a set of
orbits from points on the zero-velocity curve equally spaced in $\sin
(\beta)$ and more finely spaced than the angular bins.  The models
described here have 20 radial bins and 5 angular bins, hence our {\em
  nominal} orbit library contains $ N_{orb} = 2 \times n_r^2/2 \times
f_\beta \times n_\beta = 2 \times 20^2/2 \times 25 =$10,000 orbits.
The initial factor of 2 comes from including each orbit's retrograde
mirror image in the library, and the trailing factor of 25 results
from oversampling the 5 angular bins by a factor of $f_\beta = 5$.

Even if one does not trust this estimate, it is straightforward to
investigate whether the orbit library is large enough
by changing the number of orbits in the library. If the results are independent
of the number of orbits then presumably the library is large enough.

\vskip 10pt
\psfig{file=richstone.fig1.ps,width=9cm,angle=0} 
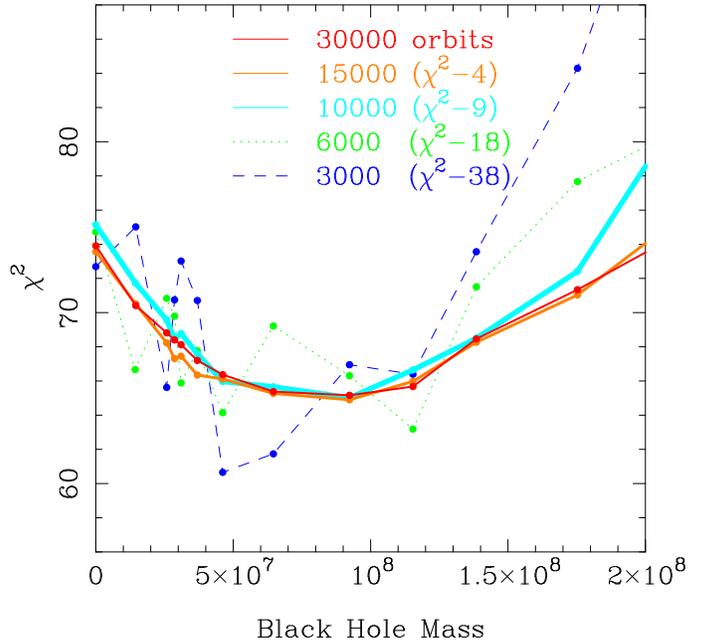
\figcaption{The
$\chi^2$ goodness of fit obtained by comparing model kinematics to
observed kinematic data for NGC 891, versus model black hole mass.
The models were constructed with successively finer coverage of phase
space (and therefore larger numbers of orbits in the library), as
described in the text.  Once the number of orbits (in this example)
exceeds 10,000 the $\chi^2$ profile is independent of the size of the
library, except for small vertical offsets which we have removed for
plotting purposes. In all these experiments we used the kinematic
dataset described in \S 3.  }
\vskip 10pt

We created an orbit library as described above to model the galaxy NGC
821 and illustrate the results in Figure 1.  The observations are
described in \citet{pink03} and in \S 3 and the modeling is described
in \citet{geb03} We chose NGC 821 to correct an error in the models in
\citet{geb03}, which used an incorrect Hubble Space Telescope (HST)
point-spread function that did not reflect the spatial binning of the
STIS CCD (this was the only one of the 12 galaxies in that paper in
which this error was made).  The revised mass is $(8.5 \pm 3.5) \times
10^7 \Msun$, at a distance of 24.1 megaparsecs, a factor of 2.3 higher than
was reported previously (at the same distance).  In order to conduct
the experiment illustrated in Figure 1 we oversampled the grid by a
factor of 2 relative to the prescription above in each dimension and
then dropped out orbits at random to obtain the libraries illustrated.
In Figure 1 we show the $\chi^2$ profile as a function of BH mass (we
have already marginalized over the galaxy mass-to-light ratio).

For small orbit libraries, Figure 1 exhibits the ragged $\chi^2$
emphasized by VME.  It also shows that for our nominal library with
$N_{orb}=$10,000 orbits, the $\chi^2$ profile is smooth enough to
infer BH mass estimates and uncertainties, and that that larger
libraries do not alter the mass estimate.  Once the number of orbits
is large enough that the $\chi^2$ profile does not change as it is
increased further, {\em the inferred BH mass does not depend on the
number of orbits}, and the phase-space coverage {\em must} be good
enough.

\vfill\eject
\section{More is Better: Data Sufficiency}

A second major point made by VME is that estimates of BH mass are
unreliable if the BH radius of influence $r_i$ is not
``well-resolved'' by the kinematic data.  This is one facet of the
rich question of what different kinds of kinematic data reveal about
the gravitational field of the galaxy.  The geometry of the kinematic
data can be characterized by: (1) the spatial resolution compared to
$r_i$; (2) the radial extent of the data (how far out the data
extend); (3) the angular coverage (how many position angles, or
whether there is integral-field data); and (4) the sparseness of the
data (in radius and angle). In addition to these geometrical
characteristics, the quality of the data (signal-to-noise and
systematic errors such as template mismatches) also determines what
one can measure.  A complete investigation of all these issues is
beyond the scope of this paper.  We emphasize the distinction between
{\em precision} (are the error bars small?) and {\em reliability} (are
the error bars accurately estimated?).  We shall argue that geometric
limitations to the quality of the kinematic data of the four kinds
listed above reduce the precision of estimates of $\mbh$, but the
range of acceptable masses can still be judged from the $\chi^2$
profile, and therefore the mass determination is reliable.

We parameterize the resolution of the kinematic data near the center
in terms of $\aleph = r_i/\theta D$, where $r_i$ is the BH radius of
influence defined earlier, $D$ is the distance, and $\theta$ is the
full-width-half-maximum telescope resolution.  VME argue that the
resolution $\aleph$ must be much greater than unity for accurate BH
mass determinations, although they do not give a clear numerical
criterion.  A number of the BH mass determinations have $\aleph$ only
slightly larger than unity. In the example of NGC 821 discussed below,
$\theta \sim 0 \farcs 08$ (the observations were made at 8500\AA\ with
a 0 \farcs 1 slit), $D=24.1 pc$ and $r_i \sim 8 pc$, so $\aleph \sim 0.9$.

VME's argument that observations with this resolution cannot determine
$\mbh$ is made without regard to the signal-to-noise of the data
available.  Even in the limit $\aleph = 0$, excellent S/N data can
reveal the presence of a BH.  Given a model with constant
mass-to-light ratio $\Upsilon$, a central mass $\mbh$ and a perfect
LOSVD measured with $\aleph \ll 1$, the shape of the LOSVD at low
velocities (comparable to velocity dispersion of the galaxy as a
whole) determines $\Upsilon$, and the shape and extent of the
high-velocity wings of the LOSVD determines the mass of a central BH.

We have investigated the effect of varying the resolution $\aleph$ on our
determinations of $\mbh$ by comparing our masses determined using both
high-resolution HST data and low-resolution ground-based data to
masses determined using the same method with ground-based data
alone. The addition of HST data generally improves $\aleph$ by a
factor of 5, from less than unity to greater than unity. The results
of these experiments are shown in Figure 8 in \citet{geb03} and Figure
4 in \citet{korm04}. We find that improved resolution generally
improves the precision of the measured $\mbh$ by narrowing the
$\chi^2$ profile, but the improved best-fit $\mbh$ always lies within
the range given by the estimated errors from the low-resolution data.
This experiment is evidence that the $\chi^2$ profiles do exactly what
they are supposed to---they provide an estimate of $\mbh$ and of its
uncertainty.  Lower resolution data yield less precise values of
$\mbh$, but not unreliable ones.

We next investigate the influence of kinematic data at larger radii on
the precision of the BH measurement (points 2 and 3 above), using data
from NGC 821 (see Figure 2). This example 
\footnote{The complete set of observations modeled in Figure 2
consists of STIS spectra and ground-based data obtained at MDM
Observatory.  The STIS data were observed through the 0\farcs 1 wide
slit with the central pixel centered on the galaxy and the data
aggregated in a set of ``observations'' binned in rectangles with
outer edges at 0\farcs 025, 0\farcs 075, 0\farcs 125, 0\farcs 225,
0\farcs 425, 0\farcs 625, 0\farcs 925.  The ground observations were
obtained through a 1\farcs 0 slit binned into rectangles with outer
edges at 1\farcs 48, 2\farcs 2, 3\farcs 6, 5\farcs 8, 9\farcs 2,
14\farcs 6, 23\farcs 2, 36\farcs 2, 51\farcs0.  The minor axis data
only extend to 23 \farcs 2.}
shows that the $\chi^2$ profile broadens and flattens as data at large
radii and along the minor axis are discarded.  The changes in the $\chi^2$
profile imply a decrease in the precision of the measurement
of $\mbh$ and indicate the increased errors.  If only
HST data is used then models with and without a BH are both acceptable,
because of the degeneracy between $\mbh$ and
$\Upsilon$ when only data inside and near the radius of influence
are used. Note that the one model with only major axis data (the
dashed line) yields a flat-bottomed $\chi^2$ profile and a very uncertain
estimate of $\mbh$ compared to the model with additional data along the
minor axis.  We conclude that inadequate spatial
coverage of kinematics far from the BH can lead to a poorly defined BH
mass and a flat-bottomed $\chi^2$ just as readily as can poorly
resolved data ($\aleph > 1$) in the galaxy center).  Again, inadequate spatial
coverage leads to imprecise BH mass estimates, but not unreliable ones.

\vskip 10pt
\psfig{file=richstone.fig2.ps,width=9cm,angle=0} 
\figcaption{
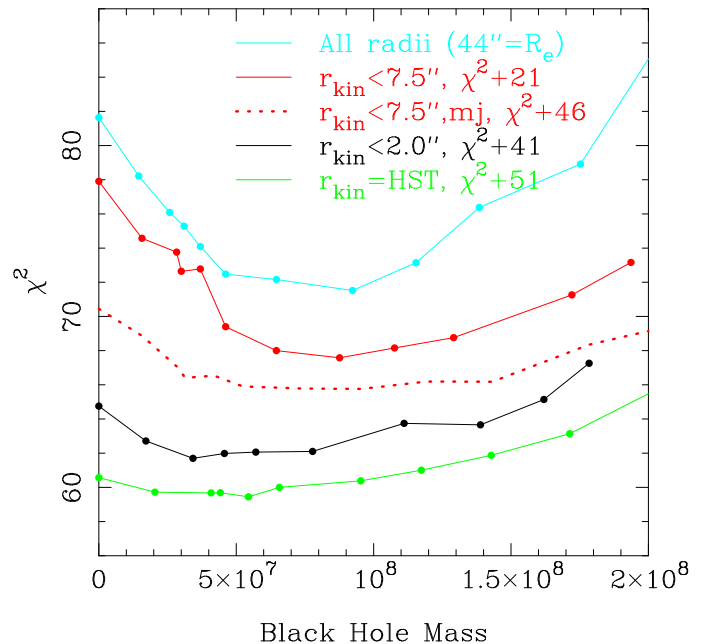
Goodness of fit $\chi^2$ versus model black-hole mass in for different
sets of observational data from NGC 821.  All models were constructed
with the 10,000 orbit library described in \S2.  All of these datasets
comprise continuous (i.e. there are no gaps in spatial coverage)
long-slit spectra from space- and ground-based observations.  The
solid lines illustrate the effect (and desirability) of extending the
najor-axis data to larger radii.  The dotted profile shows the effect
of removing minor-axis data.  If the errors in $\mbh$ are determined
by the criterion $\Delta \chi^2 = 1$ then all of the profile minima
are {\it consistent} with $\mbh = 8.5 \times 10^7 \Msun$.  }
\vskip 10pt

Turning to point 4, sparse radial sampling of the kinematic data (which
we never do) can also produce a flat bottomed $\chi^2$ profile [as
illustrated in Figure 3 in \citet{rt85}] by pushing up or down
velocity moments at unobserved radii to maximize or minimize model
parameters.

Kinematic data with better geometry (both resolution and spatial
coverage) leads to smaller uncertainties in the BH mass. However, in
all of the examples discussed above, the shape of the $\chi^2$ profile
provides an accurate estimate of this uncertainty, no matter how poor
the data geometry may be. In the absence of good spatial coverage,
data that resolves the radius of influence, or sufficiently high
signal to noise, the BH mass will simply not be well determined, but
in our experiments the true mass still lies within our error bounds.

\section{Discussion}

We have made a point of distinguishing between {\em imprecise} (the
error bars are large) measurements of the BH mass $\mbh$ and {\em
  unreliable} ones (the error bars are wrong).

Figure 1
shows that increasing the number of orbits beyond our standard
prescription does not broaden the range of BH masses allowed by our
method.  Contrary to the assertion of VME, we must be using enough
orbits.

The issue of data sufficiency is far more complex.  In \S 3 we
illustrate (or cite) several ways that data insufficiency can lead to
imprecise $\mbh$ estimates.  We also argue that these estimates are
{\em imprecise}, not unreliable.  Probably the best evidence for this
is the fact that our $\mbh$ estimates obtained with genuine
orbit-superposition models lie within their own errors of repeat
studies with greatly improved resolution \citep{geb03}\footnote
{We have not investigated filled kinematic maps in radius and angle,
  as might be obtained with integral-field devices like SAURON.
  Especially where these data imply triaxiality (thereby violating the
  basic assumption of our axisymmetric models), there may be
  unexpected changes in $\mbh$.
}.

Flat-bottomed profiles---a special case of imprecision---can be
produced by kinematic data with limited radial or angular coverage, by
sparsely sampled data, or by
insufficient resolution. A fourth way to produce a flat bottomed
$\chi^2$ profile is the use of {\em noiseless} datasets constructed
from two-integral models \citep{cretton04}.  This procedure creates
models that match the data perfectly.  Since
axisymmetric models have more freedom than two-integral models, a
range of three integral distribution functions will---by
construction---provide an {\em exact} match to the data. Since
$\chi^2$ is bounded by zero, a flat bottom in the $\chi^2$ plot is
unavoidable.

\citet{cretton04} argue that regularization (smoothing the
distribution of orbit weights) reduces the uncertainty in $\mbh$.  Our
models use a form of regularization (maximum entropy) but only as a
numerical technique to accelerate convergence: we iterate our models
while steadily reducing the weight of the entropy to the best-fit
criterion until there is no further change in the best-fit model.
\citet{verolme02b} study models with and without regularization for
M32 and find no difference in the best fit BH mass.

This paper has examined the two most serious recent criticisms of
BH mass estimates from orbit superposition.
We conclude that the issues raised by these criticisms have not
led us to report erroneous BH masses.  There are a number of other
tests of our procedure that also indicate that our method is reliable:
\begin{enumerate}
\item our method has been successfully blind-tested against (spherical) Jaffe
 models containing BHs;
\item our method yields the same masses as the Leiden code when
  applied to the same data (for NGC 821 and M32);
\item BHs weighed by this method display the same distribution of
  residuals from the mass-velocity dispersion relation as BHs weighed via gas
  dynamics \citep{dorcarn}.
\end{enumerate}
The tests described above are posted on the Nuker Team Website
\newline (http://www.noao.edu/noao/staff/lauer/nuker.html).

The BH masses published by our group
(\citealt{geb00a,bow01,geb03,sio04}) are obtained using an adequate
number of orbits.  The plots of $\chi^2(\mbh)$ we publish provide the
reader with an accurate assessment of the precision of the results.
So far as we can tell, BH masses published from the Leiden group
(\citealt{vdm98,crvdb99,verolme02b,capp1459}) also should be reliable.

\acknowledgements
We thank the director and staff of the Carnegie Institution
Observatories for hospitality during a meeting at which this paper was
developed.  Support for proposals 8591, 9106 and 9107 was provided by
a grant from the Space Telescope Science Institute, which is operated
by the Association of Universities for Research in Astronomy, Inc,
under NASA contract NAS 5-26555.  This research was also supported by
NASA Grant NAG 5-8238.

\end{document}